# Investigating Error Injection to Enhance the Effectiveness of Mobile Text Entry Studies of Error Behaviour

**Andreas Komninos, Emma Nicol and Mark Dunlop**
Computer and Information Science, University of Strathclyde
26 Richmond St., Glasgow G1 1XH
[andreas.komninos, emma.nicol, mark.dunlop]@strath.ac.uk

**ABSTRACT**
During lab studies of text entry methods it is typical to observer very few errors in participants' typing – users tend to type very carefully in labs. This is a problem when investigating methods to support error awareness or correction as support mechanisms are not tested. We designed a novel evaluation method based around injection of errors into the user's typing stream and report two user studies on the effectiveness of this technique. Injection allowed us to observe a larger number of instances and more diverse types of error correction behaviour than would normally be possible in a single study, without having a significant impact on key input behaviour characteristics. Qualitative feedback from both studies suggests that our injection algorithm was successful in creating errors that appeared realistic to participants. The use of error injection shows promise for the investigation of error correction behaviour in text entry studies.

**ACM Classification Keywords**
H.5.2 User Interfaces: Evaluation/methodology.

**INTRODUCTION**
Text entry is still core to much interaction with most computing devices including desktops, laptops, tablets, smartphones and increasingly smart TVs. There is a long and still very active history of research on developing new text entry methods for mobile phones, tablets, tables, watches and other platforms (e.g. [13][10][15]). Much of text entry research is based around comparing mobile text entry solutions through formal user-based evaluation (e.g. [13][20][22]). The evaluation of text entry approaches is key to moving the field forward, for example to achieve the challenge of inviscid text entry [9]. However, laboratory studies tend to result in very low error rates as users focus carefully on their typing. "In-the-wild" studies of users using their devices in their normal lives give the most realistic insight into mobile usage (e.g. [5]). For text entry studies in-the-wild studies can reveal behaviour of use patterns and has shown to be usable for improving text entry performance (e.g. [2]). However, in-the-wild studies are less controlled and replicable than lab based studies, are more time consuming to run, harder to recruit for as the impact is higher on participants and more complex for practitioners to adopt as keyboards have to be "production quality" to support usage outside the lab. Finally, there can be complex ethical, security and data-protection issues in logging the day-to-day text entry data that gives the truest insight into typing behaviour.

Error correction has been shown to be particularly important on touch screens with small keys (e.g. [11]) and was seen as one of the challenges for intelligent text entry [8]. Support for error awareness and correction have also been identified as a strongly desired features in studies on older adult smartphone usage [6]. Furthermore, auto-corrections errors, when current correction mechanisms fail, are widely discussed in the press as a problem of modern mobile text entry (e.g. [16]).

In this paper we present our findings of studying a novel evaluation method using both a baseline and an enhanced keyboard design, and reflect on the ability of the different study techniques to provide insights on users' text entry correction behaviour. We present a novel laboratory study technique using random-but-realistic automatic injection of errors into users' typing as they type and not on how our keyboard performed compared to a baseline keyboard. The paper first presents the background to text entry studies. We then give an overview of the case-study keyboard that is designed to support error awareness. Structured around the reporting of two user studies, the bulk of the paper focuses on our investigations into typing with injected errors using two approaches. We conclude with recommendations on using the technique to augment traditional laboratory text entry studies to gain more insight into user behaviour in error-prone situations.

**BACKGROUND: TEXT ENTRY EVALUATION**
The text entry community has widely adopted a standard approach to studies that involves users being asked to copy or transcribe a set of fixed phrases. The time that users take and number of errors they make are used as metrics to compare text entry within a study. To allow comparison between studies, standard phrase sets are now widely used. The two most widespread are the MacKenzie and Soukoreff's original 500 short-phrases set [14] (e.g. *Have a good weekend*) and the Enron Mobile collection [21] of phrases that were written on mobiles (e.g. *Can you help me here?*). There are various other specific collections such as an SMS corpus [1] and a child oriented corpus [3]. Paek and Hsu argue for the use of transcription tasks by stating that "*It is difficult, for example, to claim that one input technique performed better than another if participants using one or the other technique just happened to produce longer, more complicated phrases*" [17]. While the approach of fixed phrase copying gives strong internal consistency, reproducibility and study homogeneity advantages, the scenario of copying phrases is clearly not representative of most mobile text entry. Furthermore, a counter argument to [17] is that longer, more complicated user input with a particular keyboard may be indicative of that keyboard better supporting that user's natural typing style than, say, a simpler keyboard. As such there has been some investigation into complementing transcription tasks with more free-form entry.

© 2016 Authors
This work was partly funded by project EP/K024647/1 (EPSRC UK)

An alternative to copy tasks is to ask users to generate text in composition tasks. Karat et al. [4] compared copying sections of a novel with composing replies to scenarios and found composition speed was 58% of that for coping. Vertanen and Kristensson [22] investigated complementing copy tasks with composition tasks by asking users to (a) reply to a message, (b) compose a message without scenario prompting and (c) compose with scenario prompting. They showed that composition tasks had an entry rate of 65-85% of the copy tasks depending on task type and that typed responses varied in length between 55% and 135% of copy tasks. They concluded that "*providing participants with a simple instruction of creating a short message in the domain of interest was successful in getting participants to quickly invent and compose text. It does not appear necessary to provide participants with a specific situation or message in order to help them invent a message.*" Thus it seems that the use of composition in text entry evaluation has potential for further exploration but that it is difficult to generate longer messages than copy tasks.

The focus on much text entry evaluation is on speed of entry, usually measured in words-per-minute (WPM) following touch-typing tradition. However, accuracy is also important. In experimental conditions users are normally asked to type "quickly but accurately" and this typically results in low error rates in final submissions. Despite the typically low in-lab error rates, it is still prudent to check the correctness of final user input during studies in case a particular keyboard is error-prone and to reduce the opportunity for participants to "game" the study tasks. For copy tasks, edit distance can be used a measure of accuracy of the final phrase [19][12] or a unified measure for combining errors in the final phase with corrected errors as typing proceeds [20]. For composition tasks correctness can be inspected manually (either by the researchers or crowd sourcing [22]), by simply counting out-of-dictionary word rates or by monitoring the input stream for text corrections [23]. Given the limitations of transcription tasks and our experience of transcriptions tasks still leading to very few typing errors we wanted to explore alternatives that focus on error awareness.

## TOUCHSCREEN DESIGN FOR ERROR CORRECTION

In order to conduct comparative text entry studies we investigated the techniques using a case study based around a keyboard that was initially designed to support older adults [7], but which has error support and awareness features that have a wider potential audience. Our experience in lab studies using older and younger participants alike, have often shown very low error rates – thus few opportunities to explore behaviour on error detection and correction strategies in mobile text entry, or usage of error support features. Hence the main motivation behind our error injection technique, which was aimed to bring out more of this type of participant behaviour than is normally achievable during copy tasks. Before discussing the injected error technique, we will describe the keyboard whose novel error support mechanisms we aimed to evaluate.

### Study keyboard: The Highlighting Keyboard

We investigated usage of an Android keyboard developed to increase awareness of text entry errors [7]. The keyboard has a standard QWERTY soft keyboard layout with word suggestions that is augmented with two primary features to support error correction (see Figure 1 left): a colour bar for feedback on accuracy after typing a word, and highlighting of words within the main composition area (outside the traditional keyboard interaction area). The word highlighting was designed to emphasize errors and to support post-typing review of entered text while the coloured feedback bar was designed to give more transient feedback during typing itself and near the input focus. When finished typing a word, the keyboard automatically checks the spelling of the last word and gives the following feedback:

- Red highlight and red bar when a word is incorrectly spelled and the system is not confident of offering a good correction (a *serious* mistake);
- Orange highlight and orange bar when a word is a slight mistake and has been autocorrected as the spell checker has high confidence, or yellow highlighting if autocorrect is disabled (a *minor* mistake);
- No highlighting and green bar for a known word.

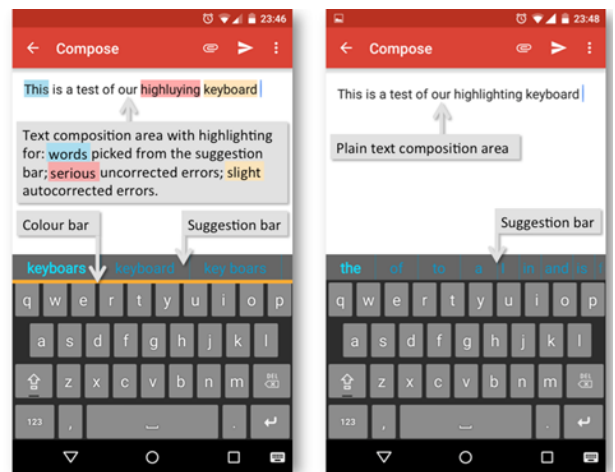

**Figure 1: Overview of the Highlighting Keyboard (left) and Normal Keyboard (right)**

The highlighting remains visible throughout the text entry sessions in order to support post-typing review. When reviewing text, tapping on a red highlight shows suggested spell corrections in the suggestion bar of the keyboard, while tapping on an orange word shows the original typed word and alternative suggestions. Auto-completions, corrections and next word suggestions are shown on a suggestion bar as standard in many soft-keyboards.

The keyboard's error-support features supplemented those already available on Android and were individually controllable through settings, allowing us to compare *Highlighting* and *Normal* study conditions (Figure 1). The *Normal keyboard* (Figure 1 right) is a simple QWERTY keyboard with a word suggestion bar; the *Highlighting keyboard* (Figure 1 left) also included the error feedback colour bar and highlighting features. Using these keyboard conditions, this paper reports on two studies into text entry behaviour, describing the evolution of our injected error algorithm.

## STUDY DESIGN AND REPORTING

In order to assess error support features of the keyboard and to get feedback from participants on its design, we ran studies with participants performing text entry tasks on the two variants of the keyboard by changing only the settings options (Figure 1 left vs right). A balanced experimental design was followed to investigate the two variants: character-by-character injection and post word completion injection (studies 1 and 2). Each study was a 2x2 within-subject design comparing the normal and highlighting keyboard variants under non-injected and injected error conditions. Different users were used for the two studies to prevent learning effects. In each study, each participant completed four input sessions by copying 14 phrases in each, in balanced order randomly assigned to participants: Normal keyboard (C1), Normal keyboard plus injected errors (C2), Highlighting keyboard (C3) and Highlighting keyboard plus injected errors (C4). We compiled 4 different phrase sets and again randomized their allocation to participants, so that the phrase sets did not correlate to the 4 keyboard conditions. The phrases in each set were selected from the Enron set [21] in a way that resulted in similar average phrase lengths for each set ($M_1$=35.3, $SD_1$=9.9, $M_2$=37.0, $SD_2$=9.5, $M_3$=38.4, $SD_3$=11.1, $M_4$=35.8, $SD_4$=7.7).

For our experiment we did not include the autocorrect option, since our desire was to look at how participants behaved when noticing errors (autocorrect obviously would have "fixed" some of the errors, hence removing the opportunity from our participants to do so themselves).

Throughout the paper we will analyse results using measures of length and speed of entry, errors while typing, and participant usage of the keyboard's suggestion bar. All data was inspected for normality using Shapiro-Wilk tests and examining the resulting Q-Q plots. Subsequent between-subject and within-subject tests were chosen according to the distribution of the relevant variables, using named parametric or non-parametric tests as appropriate. All studies were conducted under institutional ethical approval. All task sets and anonymous submitted data are available under open access terms.

## STUDY ONE: CHARACTER BY CHARACTER INJECTED ERRORS

### Version 1 – character by character error injection

Our error injection algorithm was developed over two iterations. In its first version, the algorithm was built to replace characters at the moment the user had pressed a key – simulating live errors that might be introduced by inaccurate tapping on the soft-keyboard.

In a preliminary study involving 23 users (8 aged 50+) we collected detailed typing data. Participants were asked to perform a series of copy tasks from the Enron memorable set and various composition tasks – all using the normal keyboard. Using log data from this study we developed a probabilistic model to inject errors into a user's typing in real-time as they type. In mobile text entry substitution errors tend to dominate over other types of errors due to the "fat finger" problem on mobile touch-screens [18] and this was confirmed by our logging analysis.

As we wanted to exploit all our logging data from users who typed in both copy and composition tasks, we could not rely on the "ground truth" of knowing what the user was typing to assess errors. As such we developed a heuristic of *suspect character*: a character would be considered a suspect character if it was the leftmost character deleted and replaced through a series of backspace operations, e.g. if the user typed *hrllo* followed by three backspaces and *ello* the *r* would be considered a suspect character replaced with *e*. Based on the logging of characters suspected to be input errors and their replacements, we built a 29x29 matrix (26 alphabet letters and the space, full-stop and comma characters) to represent the character substitution frequencies as derived from our logs.

|  | C1 Norm-Inject | | C2 Norm+Inject | | C3 High-Inject | | C4 High+Inject | |
| ---: | --- | --- | --- | --- | --- | --- | --- | --- |
|  | **Mean** | Std. Dev. | **Mean** | Std. Dev. | **Mean** | Std. Dev. | **Mean** | Std. Dev. |
| **Backspace ratio** | **0.111** | 0.550 | **0.144** | 0.035 | **0.108** | 0.039 | **0.139** | 0.034 |
| **Suspect key ratio** | **0.048** | 0.022 | **0.065** | 0.024 | **0.048** | 0.020 | **0.063** | 0.019 |
| **WPM** | **21.3** | 8.41 | **18.7** | 7.4 | **21.3** | 8.3 | **18.3** | 6.5 |
| **Inter-key Time (ms)** | **512** | 382 | **588** | 451 | **531** | 417 | **572** | 410 |
| **Picked Suggestions** | **15.1** | 28.5 | **19.6** | 33.2 | **15.2** | 28.3 | **18.1** | 28.6 |
| **Serious Errors** | **0.70** | 1.03 | **1.20** | 1.99 | **0.65** | 0.88 | **0.95** | 1.47 |
| **Minor Errors** | **2.30** | 3.90 | **2.60** | 5.40 | **3.85** | 8.13 | **3.85** | 7.34 |

**Table 1: Character-by-Character Injected Error Study Results Data**

To model erroneous tapping, each keystroke was given a probability threshold $P_{(t)} = 15\%$ of being selected as a candidate for a substitution error, in order to limit the total injection of errors to a realistic level. The threshold value was determined by examining the dataset and noticing that 85% of all input sessions had between 0 - 14.6% of their total input characters marked as "error suspects".

If selected for substitution the letter was replaced with a neighbouring letter with a probability based on our matrix, e.g. the 's' key would most likely be substituted with 'a' or 'd' followed by other neighbours. To enhance the realism and provide a "fat-finger" effect, we only considered the probability of keys belonging to the two-dimensional set $S$ of all characters within a 2 key width distance of the target and their substitution frequencies derived from our original matrix, to reduce the risk of wrongly modelling non-substitution errors (e.g. other spelling mistakes).

To achieve this, we build a list of substitutes as follows: If a neighbouring character $C_i$ has a substitution frequency $F_{(i)}$ of zero according to our matrix, we remove the character from S and replace it with the candidate character $C_s$ (thus creating the set $S'$). In S' we set the frequency of $C_s$ equal to the number of characters with zero substitution frequency. As an example, if a substitution candidate character $C_s$ has the neighbouring characters set $S = \{C_a, C_b, C_c, C_d, C_e\}$ with substitution frequencies $\{F_{(a)}=0, F_{(b)}=1, F_{(c)}=0, F_{(d)}=3, F_{(e)}=1\}$, the substitution set $S'$ will contain $\{C_s, C_b, C_d, C_e\}$ with frequencies $\{F_{(s)}=2$ (because $C_a$ and $C_c$ have a frequency of zero), $F_{(b)}=1, F_{(d)}=3, F_{(e)}=1\}$.

To better simulate a proximity error, we factor into the probabilities of $S'$ the distance of each replacement character key from the key of the character to be replaced, and enforce a threshold to ensure that only characters up to 2 keys apart are given a chance for replacement. To achieve this, we construct a model of the keyboard keys, taking into account their on-display sizes. We compute the distance between all the substitutes and the character to be replaced in units of "standard key distance" ($k$). Because our keys are rectangular, the $k$ metric is computed by halving the sum of a standard alphabetic letter key width and height (hence, it is between the width and the height of a typical letter key). To calculate the distance $D_{(C_i, C_j)}$ thus of two keys in terms of "standard key sizes" we compute the Euclidian distance $E_{(C_i, C_j)}$ of the two key centroids' x and y coordinates and divide this by $k$. We then divide the probabilities in $S'$ by each substitute's k value. Any keys that are more than 2 x $k$ apart, are given an large $D$ value (100 x $k$), hence effectively eliminating their chances of being picked.

The resulting probability $P_{(i)}$ thus of a character $C_i$ belonging to the set $S'$ to replace character $C_s$ thus becoming a substitute for the original character is hence defined as

$$P_{(i)} = P_{(t)} \times \frac{F_{(i)}}{\sum_{n \in S'} (F_{(n)}/D_{(n,s)})}$$

where $F_{(i)}$ is the frequency in $S'$ of the substitute character $C_i$.

**Study Participants and Procedure**

We recruited 21 participants (18 m, 3 f; age range: 17-68, mean 34.4). All participants were regular users of smartphones with touchscreen keyboards. In this experiment we enabled the autocorrect option for participants – as will be seen later, this was inconsequential, since our participants were very good at detecting mistakes as they occurred (whether naturally or injected) and correcting them on the spot.

**Results**

Results are summarized in table 1.

*Speed of typing*

The WPM rate was significantly lower during the injected error conditions for both keyboards (Normal t(20)=2.89, p=0.009; Highlighting t(20)=2.95, p=0.007, t-test). Looking in more detail at typing behaviour we also saw that the injection condition caused our participants to spend longer between each keystroke (Normal Z=-2.54, p=0.011; Highlighting Z=-2.88, p=0.004, Wilcoxon).

*Errors while typing*

In Table 1, we report the average ratios of backspaces and suspect characters per total typed characters. Injection of errors led to significantly higher backspace use and an increased frequency of suspect characters during use of both keyboards (Backspaces: Normal t(20)=3.30, p0.004; Highlighting t(20)=4.05, p<0.001. Suspects: Normal

t(20)=-3.63, p=0.002; Highlighting t(20)=5.20, p<0.001, t-tests).

We analysed errors based on words being committed to the text area from the keyboard – normally by the user typing space or punctuation after the word. Table 1 shows the average serious and minor errors per sentence. Although mean error rates were higher for injected conditions, neither the mean increase in serious nor minor errors was significant (Serious: Normal Z=-0.86, p=0.390; Highlighting Z=-1.24, p=0.215. Minor: Normal Z=-0.18, p=0.857; Highlighting Z=-0.68, p=0.497, Wilcoxon). Likewise the highlight keyboard did appear to result in more minor errors but this was also not significant.

*Suggestion bar usage*
Participants preferred to do most of their input character-by-character and made little use of the suggestion bar with no difference between conditions (Normal Z=-1.33, p=0.184; Highlighting Z=-0.80, p=0.424, Wilcoxon). Table 1 shows the average number of times a suggestion was picked per sentence. An interesting observation, however, is that almost all usage of the suggestion bar came from our older participants (over 50s), while younger participants almost completely ignored it.

*Subjective Feedback*
Analysis of NASA TLX data showed no significant differences for Mental, Physical, Temporal or Performance metrics. However, in the case of Overall Effort and Frustration levels the injection did lead to a difference between the keyboard conditions: the normal keyboard showed no significant difference with and without injection, while the highlighting keyboard showed significantly higher effort (Z=-2.80, p=0.005, Wilcoxon) and frustration (Z=-2.86, p=0.004, Wilcoxon).

|  | C1 Norm -Inject | C2 Norm +Inject | C3 High -Inject | C4 High +Inject |
|---|---|---|---|---|
| **Mental** | 10.0 | 9.0 | 7.5 | 7.5 |
| **Physical** | 6.0 | 6.0 | 7.0 | 6.5 |
| **Temporal** | 10.0 | 10.0 | 10.0 | 10.0 |
| **Performance** | 8.0 | 9.5 | 7.0 | 9.0 |
| **Effort** | 11.5 | 13.5 | 10.0 | 12.5 |
| **Frustration** | 10.0 | 11.0 | 9.0 | 12.0 |

**Table 2 NASA TLX Median Scores**

In a post-study questionnaire participants reported that they did indeed notice injections as they happened (in response to "I always noticed the injected errors as I was typing" users answered with a median of 4 [agree] on a 5 point Likert scale). This is backed by self-reporting of more careful behaviour while typing (median 4 [agree] to both "Injected errors made me type more carefully" and "The injected errors changed how I typed"). However participants also supported the "realism" of the mistakes being injected into the text ("The injected errors were similar to errors I make when typing quickly on my phone" and "The injected errors were similar to errors I make when typing on my phone while on the move" both median 4). Where they disagreed was with the frequency of injected error appearance, they thought that this differs from what they would observe in real life (median 2 [disagree] to both "The frequency of injected errors was similar to the frequency of errors I make while typing quickly" and "The frequency of injected errors was similar to the frequency of errors I make while typing on the move"). In discussion it was clear they felt we were injecting somewhat more, but not excessively more, errors.

**Discussion and Limitations**
Taken together the results on increased backspace usage, longer inter-key times and no significant difference in error insertion strongly indicates that users simply slowed their typing to correct injections as they typed rather than correcting words after completion.

Although there's some evidence of increased errors by some users, overall users simply slowed down to keep entry accurately character-by-character. As a result our error supporting mechanisms were still not being exercised and we were not noticeably increasing error rates. A pattern is emerging that in our lab studies users react to test situations by typing as slowly as necessary to maintain very high typing quality.

**STUDY TWO: WORD BY WORD INJECTED ERRORS**
Character-by-character injection succeeded in introducing errors that were, overall, considered to be realistic by users. However, the method had too strong an effect on typing behaviour with users slowing down typing considerably to compensate. As an alternative we investigated injecting errors after each word was completed rather than on a key-by-key basis and introduced version 2 of our injected errors algorithm. This variant is similar to auto-correction replacing words with unintended words when the word is completed, except in our case the resulting word is unlikely to be a valid dictionary word.

Following the results of study 1, we amended the algorithm so that it would only effect error injections after the user had finished typing in a word. Because the algorithm is agnostic to the length of the input word, there are possibilities (although small) that more than one characters in the same word may be replaced, resulting in final input that seems unlikely (e.g. because the characters 't' and 'r' have a high likelihood of being replaced with 'y' and 'e' respectively, a user may have typed in the word 'try' only to see it being replaced with 'yey'- a level of error injection we felt would be excessively noticeable). To prevent such problems, we limit the number of allowed replacements within any given word w to [1, (0.25 x $L_{(w)}$)] where L(w) is the word length. If the number of replacements exceeds the replacement cap, then we reduce the number of replacements down to the upper limit by removing as many replacements as required in a random manner. The cap is enforced only after the user has finished typing a word and the errors are injected into the completed word, replacing the characters the user has already typed in.

|  | C1 Norm – Inject | | C2 – Norm + Inject | | C3 High – Inject | | C4 High + Inject | |
|---|---|---|---|---|---|---|---|---|
|  | **Mean** | Std. Dev. | **Mean** | Std. Dev. | **Mean** | Std. Dev. | **Mean** | Std. Dev. |
| **Backspace ratio** | **0.06** | 0.03 | **0.13** | 0.03 | **0.08** | 0.03 | **0.13** | 0.04 |
| **Suspect key ratio** | **0.04** | 0.02 | **0.05** | 0.02 | **0.04** | 0.02 | **0.05** | 0.02 |
| **WPM** | 21.94 | 3.54 | 18.27 | 2.11 | 21.5 | 4.54 | 17.72 | 2.62 |
| **Interkey time (ms)** | 421 | 127 | 505 | 122 | 457 | 118 | 484 | 99 |
| **Picked suggestions** | 0.99 | 1.03 | **1.00** | 1.27 | **0.75** | 1.08 | **0.79** | 0.76 |
| **Serious errors** | 0.14 | 0.08 | **0.30** | 0.15 | **0.15** | 0.10 | **0.34** | 0.13 |
| **Minor errors** | 0.34 | 0.27 | **1.51** | 0.55 | **0.45** | 0.26 | **1.53** | 0.33 |
| **Input accuracy** | 0.19/0.20 | 0.26/0.35 | **0.40/0.32** | 0.40/0.30 | 0.11/0.10 | 0.14/0.11 | **0.38/0.33** | 0.38/0.39 |

**Table 3: Word-by-Word Completion Injected Error Study Results Data**

Given this process, we handle the deletion of characters during the composition of a word as follows: Say for example that the user typed in the word "toying" and as they get to the final character, and before pressing a word terminator such as space, full stop, comma etc., decide to change the word to "toyed". Also let's assume in this example that the algorithm determined that the character 'o' will be replaced by a neighbouring character, e.g. 'i', and that the character 'n' will be replaced by 'b', when the user has finished typing. As the user starts to backspace, the algorithm will keep the 'i for o' substitution, as this precedes the characters that have been deleted, and the 'b for n' substitution will be discarded. As the user continues to type the characters following the sequence 'toy-', the algorithm will calculate the replacement probabilities for any new characters being input.

**Study Participants and Procedure**

We recruited 20 participants (13 m, 7 f; age range: 22-33, mean 26.05). Again all participants were regular users of smartphones with touchscreen keyboards. Our aim was to discover whether the modified injection algorithm encourages participant behaviour during input and whether it provides more opportunities to study participant behaviour in handling errors during input. Our aim was that the injection algorithm should not significantly affect the core properties of participant behaviour during typing, but should provide more opportunities to investigate behaviours and strategies when noticing errors. To this end, we performed a second study using the same 2x2 design (keyboards and injection).

**Results**

Results are summarized in Table 3 and Figure 2.

*Speed of typing*

With regard to the WPM rate, in both keyboards participants exhibited a statistically significant lower WPM rate in the injected condition (Normal t(19)=4.813, p<0.001 t-test, Highlighted Z=-3.136, p=0.002 Wilcoxon). It is noteworthy however that the difference is small, approximately 3-4WPM in both cases. These findings are in line with those of the previous experiment. With regard to inter-key times, we find a statistically significant difference only in the Normal keyboard (t(19)=-3.537, p=0.002, t-test), whereas in our previous experiment, this difference was observed in both keyboards.

*Errors while typing*

In Table 3, we report the average ratios of backspaces and suspect characters per total typed characters. We observed that in both keyboards, the injected condition caused heavier usage of the backspace key, as was expected. This difference was statistically significant both in the Normal (Z=-3.883, p<0.001, Wilcoxon) and Highlighted keyboards (t$_{(18)}$=-3.963, p<0.001). This finding contrasts the results of our previous experiment, where the participants were able to notice injected errors as they occurred. We noted also that the suspect character ratio was statistically significant only for the Normal keyboard (Z=-3.360, p<0.001), compared to our previous experiment where this difference was statistically significant in both. It is interesting here to note that the injected conditions increased the number of backspaces disproportionately to the number of resulting suspect characters, which indicates that in the injected conditions, errors were spotted at places nearer the beginning of a word, hence requiring more backspaces to fix.

Regarding minor and serious mistakes, Table 3 shows the average serious and minor errors per sentence. The injected condition clearly caused more to appear in both keyboards (Minor: Normal Z=-3.922, p<0.001, Wilcoxon, Highlighted: t(19)=-12.944, p<0.001, t-test, Serious: Normal Z=-3.469, p<0.001, Wilcoxon, Highlighted Z=-3.433, p<0.001 Wilcoxon). This was an expected result from the way our algorithm was modified to behave.

*Suggestion bar usage*

In terms of suggestion picking behaviour, we did not discover any statistically significant differences in the

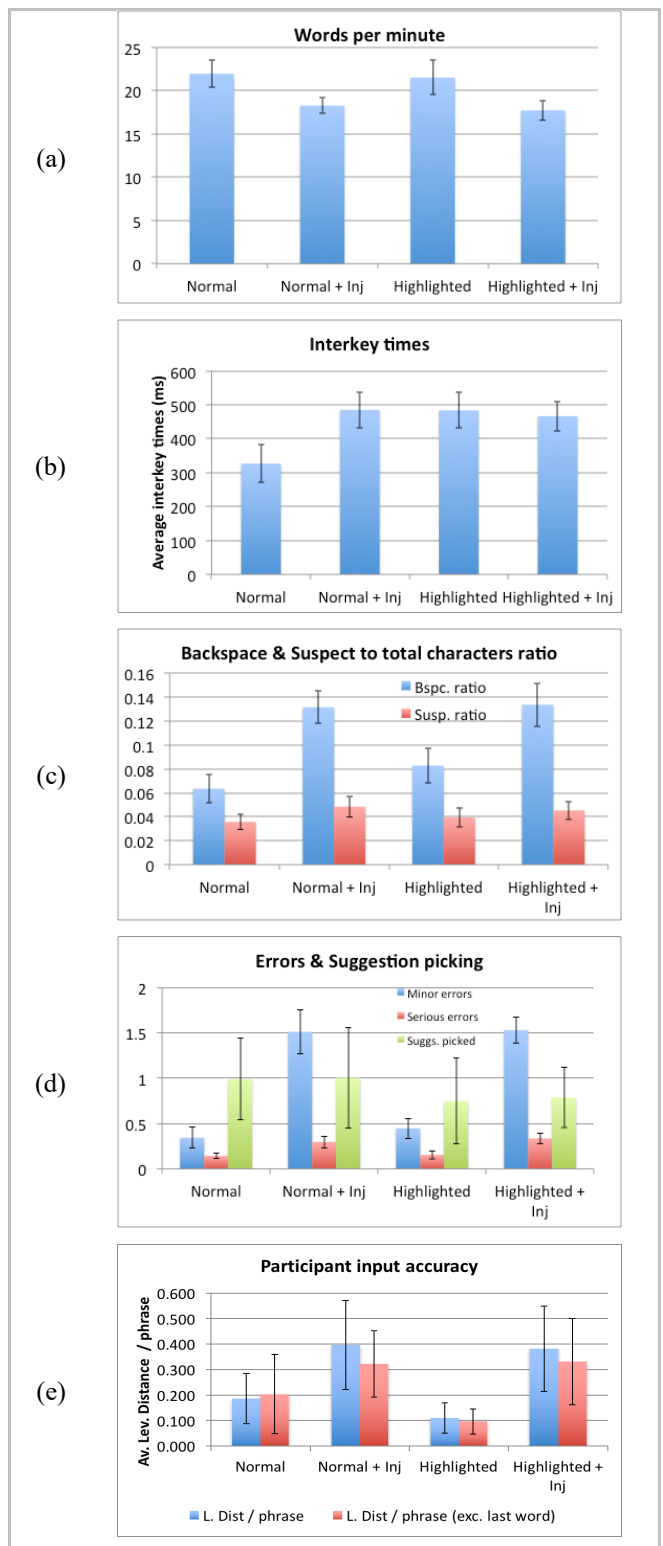

**Figure 2. Participant behaviour in study 2.**

*Input accuracy*
Given that our algorithm successfully caused the emergence of significantly more serious and minor mistakes, we were able in this case to discover whether our participants were able to detect these and make the necessary corrections before submitting their text. Table 3 shows the average Levenshtein distance per sentence (calculated by converting to lower case and trimming punctuation and trailing spaces) for the entire submitted sentence and with the last word removed (see below for explanation).

In both the Normal and Highlighted keyboards, we notice that the differences in participants' average Levenshtein string distance between submitted and requested text was higher with statistical significance in the injected conditions (Normal: Z=-2.272, p=0.023, Wilcoxon, Highlighted: Z=-2.850, p=0.004, Wilcoxon). This result shows that participants' input accuracy suffered due to the injection of errors, of which a larger percentage went undetected during submission when the injection condition was enabled. Here we were somewhat concerned that our Highlighted keyboard's error detection support mechanisms did not allow participants to detect the mistakes and correct them. We noted that most of the uncorrected mistakes occurred at the end of input, for participants that forgot to press the full-stop character at the end and immediately pressing the "submit" button. The session ending caused injections in the last word to be affected, without giving our participants a chance to correct the mistake. Further analysis with the removal of the last word from each submission led to slightly reduced means for average distance (Normal: Z = -1.835, p=0.066, Highlighted: Z= -2.388, p<0.05, Wilcoxon) with little effect on non-injected conditions (Figure 2e).

*Subjective Feedback*
In Study 2, the NASA TLX results (Table 4) showed that injection caused a higher mental demand in the Normal keyboard (Z=-2.472, p=0.013, Wilcoxon) but not the highlighted keyboard. Additionally, participants felt their performance was worsened during the injected condition with the Normal keyboard (Z=2.400, p=0.016, Wilcoxon) but not the Highlighted keyboard. Effort was again greater in the Normal keyboard with the injected condition (Z=-3.131, p=0.002, Wilcoxon) but not the Highlighted keyboard. Frustration however was greater in both keyboards with the injected condition (Normal: Z=-3.731, p<0.001, Highlighted: Z=-2.490, p=0.013, Wilcoxon). Physical and Temporal demand did not exhibit any statistically significant differences due to the injected condition, in both keyboards.

means, for both keyboards. Table 3 shows the average number of times a suggestion was picked per sentence.

| | C1<br>Norm<br>-Inject | C2<br>Norm<br>+Inject | C3<br>High<br>-Inject | C4<br>High<br>+Inject |
|---|---|---|---|---|
| Mental | 10.0 | 9.0 | 8.0 | 9.0 |
| Physical | 7.0 | 6.0 | 7.0 | 7.0 |
| Temporal | 10.0 | 10.0 | 10.0 | 10.0 |
| Performance | 9.0 | 10.0 | 7.5 | 9.5 |
| Effort | 11.5 | 13.0 | 10.5 | 12.0 |
| Frustration | 10.0 | 11.0 | 9.0 | 12.0 |

Table 4: Study 2 NASA TLX Medians

**Studying participant behaviour during errors**

Given that our injection algorithm caused the appearance of more serious and minor mistakes in both keyboards, our next goal was to determine whether this provides more opportunities to observe participant behaviour strategies in correcting mistakes when they are spotted. One such metric of correction strategies is to look at the backspacing sequences employed by participants. We note here that our keyboard does not support a "long-press" of the backspace key (which in some keyboards results in deletion of an entire word), hence each backspace press (even long ones) deletes exactly one character. By looking at the instances of consecutive backspace presses and the length of these sequences, we observe that in both keyboards, the injected condition causes a higher frequency of longer backspace sequences. In the non-injected conditions, 90% of all backspacing sequences are up to 3 (Normal) or 4 (Highlighted) backspaces long. In contrast, in injection conditions, the 90% cut-off is at 6 consecutive backspaces, indicating clearly that our algorithm provides more opportunities to examine participant behaviour when mistakes are detected only after a word has been typed, or when they are early into a relatively long word (e.g. forcing participants to make a choice on whether to consecutively backspace, hence possibly missing their target, or to attempt to move the cursor at or near the position of the mistake).

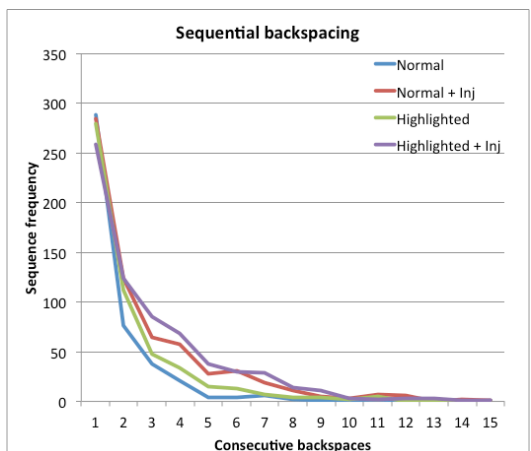

Figure 3. Frequency of employing the sequential backspacing correction strategy

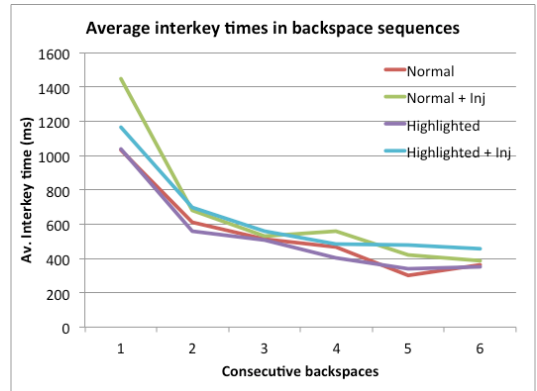

Figure 4. Inter-key times in single and consecutive backspaces

By electing to move the cursor to a previous position in the text instead of consecutively backspacing, a user postpones interaction with the keyboard keys in order to have to press a small number of backspaces (one or two, if they have accurately positioned the cursor near the mistake). Effects of such behaviour are noticeable in our results, if we consider the average inter-key time in each sequence of backspaces. We can clearly see that single backspace sequences exhibit a longer average inter-key time, which drops dramatically for double, and then continues to drop for multiple consecutive backspaces. We can see that in the injected conditions, these inter-key times (particularly for 1 and 2 backspaces) are longer, showing that participants took longer to move the cursor (i.e. the spotted mistake was earlier in the text) than in non-injected conditions, hence the injection algorithm affords researchers opportunities to study the movement of cursors to detect mistakes as a strategy.

We notice also that our algorithm provides ample opportunity to study behaviour during extreme backspacing that spans multiple words (e.g. 51 and 62 cases of 7-12 consecutive backspaces in the injected condition Normal & Highlight keyboards respectively, compared to just 12 and 22 in the non-injected condition).

**CONCLUSION**

In previous studies we noticed that participants in a lab environment were especially careful during input, correcting mistakes as they occurred and checking each letter as it was being type. This provided very little opportunity to study their behaviour in correcting mistakes that had gone unnoticed, made it difficult to try out novel mechanisms for assisting in the spotting of mistakes as they occur during input. In contrast with participants' expressed concerns over undetected errors in text, particularly when sent to another person, along with the commonplace understanding of how embarrassing auto-correct can be when it replaces words out of context to the intended message, we were unable to reproduce enough behaviour of this type in the constraints of lab studies without involving either a large number of participants, or assigning participants a significantly larger workload in terms of input tasks. In Study 1, we attempted to artificially overcome participants' careful behaviour by injecting artificial but realistic errors in their input, as it took place. Despite overall feeling that the errors were representative, taken together the results on increased backspace usage, longer inter-key times and no significant difference in error insertion strongly indicates that users simply slowed their typing to correct injections as they typed rather than correcting words after completion.

In contrast, Study 2 was much more successful in bringing out a wider diversity of error management strategies by users. We saw during the use of both keyboards clear evidence of different patterns during backspace usage, including an increased frequency of longer backspace sequences and evidence of strategic decisions on positioning the cursor near a mistake, rather than just backspacing. Furthermore, participants did not perceive that injected errors caused a greater temporal demand in either case.

As stated in the introduction, the purpose of this paper was not to examine whether the highlighting method offers any advantages in detecting and fixing errors, compared to a plain QWERTY keyboard. Our purpose was to test whether the injected algorithm can provide more opportunities for researchers to study error management behaviour in a lab setting, without requiring great numbers of participants or burdening the participants with an excessive number of tasks to complete. To this end we feel that our algorithm has shown promise in enticing a greater frequency and diversity of error correction behaviours, which deserve further research. As such we recommend use of word-by-word error injection in studies that wish to investigate user correction behaviour or support mechanisms.

We have noticed in our experiments that participants frequently "context-switch", i.e. shift the focus of their attention between the keyboard and the input area, to check for mistakes. The design of multimodal feedback cues to remove the need for such frequent context-switches may yield performance improvements that can be tested using lab tasks and our algorithm. The algorithm itself can be improved to position the injected errors further from the current focus of the user's attention. For example, instead of modifying the word that has just been composed, the algorithm could be easily made to modify characters in the previous two or three words, hence making it more difficult for the user to immediately notice any changes But at a further cost in realism. A hybrid approach of infrequently injecting character-by-character errors along with word-by-word errors may also prove productive and be more realistic. Other improvements may include a learning, adaptive strategy which will be trained by each individual user's own mistakes and thus produce errors which are more realistic to each individual user. Our *suspect character* heuristic would permit in-the-wild collection of data in a manner that does not require recording of full text-entry and thus could avoid some privacy concerns.

At this time we are continuing our research and hope to be able to report on the design of new feedback and support mechanisms for mobile text entry, as well as modifications and improvements to our method.

The source code for our keyboard and all data described in this paper are at
https://pureportal.strath.ac.uk/en/projects/empirical-investigation-user-centred-development-of-touch-screen-